\documentclass[aps,prb,showpacs,groupedaddress,superscriptaddress,
twocolumn]
{revtex4}
\usepackage{graphicx,bm}

\begin{document}

\title{Strong magnetic pair breaking in Mn substituted MgB$_2$ single crystals}

\author{K. Rogacki}
\affiliation {Laboratory for Solid State Physics, ETH Z\"{u}rich,
8093 Z\"{u}rich, Switzerland} \affiliation {Institute of Low
Temperature and Structure Research, Polish Academy of Sciences,
50-950 Wroclaw, P.O.Box 1410, Poland}

\author{B. Batlogg}
\affiliation {Laboratory for Solid State Physics, ETH Z\"{u}rich,
8093 Z\"{u}rich, Switzerland}

\author{J. Karpinski}
\affiliation {Laboratory for Solid State Physics, ETH Z\"{u}rich,
8093 Z\"{u}rich, Switzerland}

\author{N. D. Zhigadlo}
\affiliation {Laboratory for Solid State Physics, ETH Z\"{u}rich,
8093 Z\"{u}rich, Switzerland}

\author{G. Schuck}
\affiliation {Laboratory for Solid State Physics, ETH Z\"{u}rich,
8093 Z\"{u}rich, Switzerland}

\author{S. M. Kazakov}
\affiliation {Laboratory for Solid State Physics, ETH Z\"{u}rich,
8093 Z\"{u}rich, Switzerland} \affiliation {Department of
Chemistry, Moscow State University, 119899 Moscow, Russia}

\author{P. W\"{a}gli}
\affiliation {Laboratory for Solid State Physics, ETH Z\"{u}rich,
8093 Z\"{u}rich, Switzerland}

\author{R. Pu\'{z}niak}
\affiliation {Institute of Physics, Polish Academy of Sciences,
Aleja Lotnik\'{o}w 32/46, 02-668 Warsaw, Poland}

\author{A. Wi\'{s}niewski}
\affiliation {Institute of Physics, Polish Academy of Sciences,
Aleja Lotnik\'{o}w 32/46, 02-668 Warsaw, Poland}

\author{F. Carbone}
\affiliation {D\'{e}partement de Physique de la Mati\`{e}re
Condens\'{e}e, Universit\'{e}e de Gen\`{e}ve, 1211 Gen\`{e}ve 4,
Switzerland}

\author{A. Brinkman}
\affiliation {D\'{e}partement de Physique de la Mati\`{e}re
Condens\'{e}e, Universit\'{e}e de Gen\`{e}ve, 1211 Gen\`{e}ve 4,
Switzerland}

\author{D. van der Marel}
\affiliation {D\'{e}partement de Physique de la Mati\`{e}re
Condens\'{e}e, Universit\'{e}e de Gen\`{e}ve, 1211 Gen\`{e}ve 4,
Switzerland}

\date{\today}

%---------------------------------------------------------------------
\begin{abstract}

Magnetic ions (Mn) were substituted in MgB$_2$ single crystals
resulting in a strong pair-breaking effect. The superconducting
transition temperature, $T_c$, in Mg$_{1-x}$Mn$_x$B$_2$ has been
found to be rapidly suppressed at an initial rate of 10~K/$\%$Mn,
leading to a complete suppression of superconductivity at about
2$\%$~Mn substitution. This reflects the strong coupling between
the conduction electrons and the $3d$ local moments, predominantly
of magnetic character, since the nonmagnetic ion substitutions,
e.g. with Al or C, suppress $T_c$ much less effectively (e.g.
0.5~K/$\%$Al). The magnitude of the magnetic moment ($\simeq$
1.7~$\mu_B$ per Mn), derived from normal state susceptibility
measurements, uniquely identifies the Mn ions to be divalent, and
to be in the low-spin state ($S =$ 1/2). This has been found also
in $X$-ray absorption spectroscopy measurements. Isovalent
Mn$^{2+}$ substitution for Mg$^{2+}$ mainly affects
superconductivity through spin-flip scattering reducing $T_c$
rapidly and lowering the upper critical field anisotropy
$H_{c2}^{ab}/H_{c2}^c$ at $T = 0$ from 6 to 3.3 ($x = 0.88$\% Mn),
while leaving the initial slope d$H_{c2}$/d$T$ near $T_c$
unchanged for both field orientations.

\end{abstract}



\maketitle

%---------------------------------------------------------------------
\section{INTRODUCTION}

In conventional superconductors magnetic interactions and magnetic
impurities are generally destructive to
superconductivity.\cite{Maple'AP'1976, Maple'Fischer'1982}
However, when the interaction is weak and of an antiferromagnetic
(AFM) type, both magnetism and superconductivity can coexist as in
classic,\cite{Maple'Fischer'1982, Canfield'PT'1998,
Muller'RPrPh'2001} heavy fermion,\cite{Flouquet'cond-mat'0505713}
and high-temperature superconductors.\cite{Maple'PhysicaB+C'1987,
Maple'JLCM'1989, Nachtrab'cond-mat'0508044} Moreover, in some
unconventional quantum systems, such as Sr$_2$RuO$_4$
(Ref.~\onlinecite{Ishida'Nature'1998}), UGe$_2$
(Ref.~\onlinecite{Saxena'Nature'2000}), and ZrZn$_2$
(Ref.~\onlinecite{Pfleiderer'Nature'2001}), superconductivity
appears to be mediated by magnetic interactions. Within this
context, studies of the interaction between magnetic impurities
and superconductivity in the unconventional two-gap superconductor
MgB$_2$ emerge as an important task.

Shortly after the discovery of 40-K superconductivity in
MgB$_2$,\cite{Nagamatsu'Nature'2001} the intensive studies of its
electronic structure revealed that this compound is a two-gap
multi-band superconductor with two-dimensional (2D) $\sigma$-band
and three-dimensional (3D) $\pi$-band.\cite{Kortus'PRL86'2001,
Liu'PRL'2001, Choi'Nature'2002, Canfield'PT'2003} The high
superconducting transition temperature, $T_c$, is mainly
associated with the $\sigma$-band, and $T_c$ depends on both the
electron (hole) doping intensity, which changes the Fermi level
and the Fermi surface geometry,\cite{Pena'PRB'2002,
Kasinathan'PhysicaC'2005, Klie'cond-mat'0510002} and the interband
scattering,\cite{Golubov'JPCM'2002, Mitrovic'JPCM'2004,
Kortus'PRB'2005, Dolgov'PRB'2005} which may also influence the
anisotropy.\cite{Golubov'PRB'1997, Angst'PRB71'2005} On the other
hand, as in conventional superconductors, $T_c$ is expected to be
affected by the pair-braking effect caused by magnetic impurities
or substitutions that suppress superconductivity due to the
exchange interaction between conduction electrons and magnetic
moments of the substituted ions.\cite{Moca'PRB'2002} The magnetic
pair-braking effect has been studied intensively in
classic\cite{Fischer'ApplPhys'1978, Shrivastava'PhysRep'1984,
Goldman'PRB'1994, Chervenak'PRB'1995, Bill'PhysicaC'1998,
Ghosh'PRB'2001, Kim'PRB'2002} and high-temperature
superconductors,\cite{Xiao'PRB'1990, Szabo'PRB'2000,
Chattopadhyay'JPCM'2002, Poddar'EPJB'2003, Rogacki'PRB'2003}
however this effect has remained almost untouched in more exotic
superconductors, particularly in two-gap multi-band MgB$_2$, where
only two reports on such studies has been
published.\cite{Moca'PRB'2002, Dolgov'PRB'2005} The main goal of
this work is to study the influence of magnetic Mn-ion
substitutions on the normal-state and superconducting properties
of high-quality MgB$_2$ single crystals.  The results are analyzed
and discussed in the context of nonmagnetic ion substitutions that
affect superconductivity considerably less.

In conventional superconductors the substitution of a small amount
of magnetic impurities destroys superconductivity but the addition
of nonmagnetic ions is rather harmless. In unconventional
multi-band multi-gap superconductors, both magnetic and
nonmagnetic impurities may affect superconductivity in a similar
way, depending on intraband and interband scattering.  An enormous
activity in experimental and theoretical studies has been
performed to explain the puzzling behavior of the multi-band
two-gap superconductor MgB$_2$ substituted with nonmagnetic ions.
Here, the most intensively studied substitutions are Al for Mg
(Ref.~\onlinecite{Slusky'Nature'2001, Cava'PhysicaC'2003,
Putti'PRB'2003, Karpinski'PRB'2005, Zambano'SST'2005}) and C for B
(Ref.~\onlinecite{Bharathi'PhysicaC'2002, Lee'PhysicaC'2003,
Masui'PRB'2004, Kazakov'PRB'2005}); both fill the MgB$_2$
hole-bands with electrons and also introduce scattering centers
that may act in different ways. In spite of this great
experimental and theoretical effort, the superconducting and
normal-state properties of MgB$_2$ substituted with magnetic ions
have been investigated briefly and in polycrystalline
materials.\cite{Xu'JPSJ70'2001, Kuhberger'PhysicaC'2002,
Dou'SST'2005, Zhang'PhysicaC'2005} Concomitantly to this work, the
Mn-substituted MgB$_2$ crystals from the same batches have been
studied by a point-contact
spectroscopy.\cite{Gonnelli'cond-mat'0510329}

In this study we focus in detail on the crystallographic,
magnetic, and superconducting properties of Mn-substituted MgB$_2$
single crystals. We report a rapid reduction of the
superconducting transition temperature $T_c$ due to the magnetic
ion substitution and, in contrast, a moderate influence of Mn on a
temperature dependence of the upper critical field, $H_{c2}$, and
the critical field anisotropy $\gamma = H_{c2}^{ab}/H_{c2}^c$. A
central question in the discussion on the influence of Mn on
superconducting properties of MgB$_2$ is the valence state of Mn
and the spin configuration of its $d$-electrons. We have studied
the magnetic state of Mn by measuring the normal state
magnetization and the X-ray absorption involving the $3d$
electrons. All modifications of the superconducting properties are
consistent with strong magnetic pair breaking by Mn$^{2+}$ ions
with $S = 1/2$.

%---------------------------------------------------------------------
\section{EXPERIMENTAL}

Single crystals of Mg$_{1-x}$Mn$_x$B$_2$ have been grown under
high pressure using the cubic anvil press. A mixture of Mg, Mn, B,
and BN is placed in a BN crucible in a pyrophyllite cube. (For
example, the Mg:Mn:B:BN ratio of 9.5:0.5:12:1 results in crystals
with 2$\%$ of Mn substituted.) The inner diameter of the crucible
is 8~mm, and its length is 8.5~mm.  The heating element is a
graphite tube. Six anvils generate pressure on the whole assembly.
The typical growth process involves: (i) increasing of pressure up
to 30~kbar, (ii) increasing of temperature up to 1960~$^\circ$C in
1~h, (iii) dwelling for 0.5-1~h, (iv) lowering the temperature and
pressure in 1~h. As a result, Mg$_{1-x}$Mn$_x$B$_2$ crystals
sticking together with BN crystals have been obtained. Using this
method, Mg$_{1-x}$Mn$_x$B$_2$ crystals up to $0.8 \times 0.8
\times 0.1$~mm$^3$ have been grown. The phase purity of the
crystals has been confirmed by X-ray diffraction. The Mn content
has been determined by energy dispersive X-ray (EDX) analyses. For
all Mn substitutions from 0.4 to 7~$\%$, the crystals are single
phase and homogenous, at least within $\pm 0.04\%$ of Mn content.

The lattice parameters of Mn-substituted crystals were determined
by a four-circle single crystal X-ray diffractometer Siemens P4
with molybdenum $K_{\alpha1}$ radiation. A set of 32 reflections
recorded in the range of $2\Theta$ angle ($15^\circ < 2\Theta <
32^\circ$) was used to calculate the unit cell parameters.
Detailed structure analysis was performed for several
Mg$_{1-x}$Mn$_x$B$_2$ single crystals with Mn content up to $x =$
0.07. Measurements were carried out on a Bruker SMART CCD system
with molybdenum $K_{\alpha1}$ radiation. The refinement of
Mg$_{1-x}$Mn$_x$B$_2$ structure with Mn on Mg position was
successful and no phase separation was observed.\cite{Schuck'2006}

Magnetic properties in the normal and superconducting states were
investigated by magnetic moment measurements performed as a
function of temperature and field with a Quantum Design Magnetic
Properties Measurement System (QD-MPMS) equipped with a 7~T
superconducting magnet. Individual crystals with a mass of about
25~$\mu$g as well as a collection of 25 crystals with a mass of
847~$\mu$g were studied to obtain more reliable quantitative
results. In order to determine the upper critical field, the
magnetic moment $M$ was measured at constant field upon heating
from the zero-field-cooled state (ZFC mode) or the field-cooled
state (FC mode), with a temperature sweep of 0.1~K/min.
Occasionally, $M$ was also measured at constant temperature with
increasing field, using the step-by-step option. Complementary
torque measurements were performed to obtain the upper critical
field properties at higher fields. The torque $\mathbf{\tau}=
\mathbf{M} \times \mathbf{B} \approx \mathbf{M} \times \mathbf{H}$
was recorded as a function of the angle between the applied field
and the $c$-axis of the crystal for various fixed temperatures and
fields. For the torque measurements, a QD Physical Properties
Measurement System (QD-PPMS) with torque option and a maximum
field of 9~T was used. For details of torque measurements see Ref.
\onlinecite{Angst'PRL'2002}.

X-ray Absorption Spectroscopy (XAS) measurements on the $2p$ to
$3d$ absorption threshold of Mn impurities in MgB$_2$ single
crystals have been performed on the beam line BACH at "Elettra"
synchrotron (Trieste).\cite{Zangrando'RSI'2001} The spectra have
been collected both in Total Electron Yield (TEY) and Total
Fluorescence Yield (TFY) at room temperature. The TEY technique
measures the photoconductivity of the sample as a function of the
incoming photon energy, this quantity is directly related to the
optical absorption cross section. This method is sensitive to the
first 50~\AA~of the sample, which means that the contamination of
the surface could affect the shape of the spectrum. The TFY method
measures the integrated intensity of the fluorescence decay
$3d$$\rightarrow$$2p$ as a function of the incoming photon energy.
This quantity is usually not one to one related to the optical
absorption cross section because of self absorption and saturation
phenomena. These phenomena are less important when the fluorescent
ion is present at low concentrations, which makes TFY particularly
suitable for studying the absorption spectra of diluted solutions
or impurities in crystals.\cite{Groot'SSC'1994} The main advantage
of TFY is the bulk sensitivity, probing the first 200~nm of the
sample. On the other hand, TFY is experimentally more demanding,
resulting in a longer acquisition time and a poorer resolution.

%---------------------------------------------------------------------
\section{RESULTS AND DISCUSSION}

\begin{figure}[!htb]
\includegraphics*[width=0.45\textwidth]{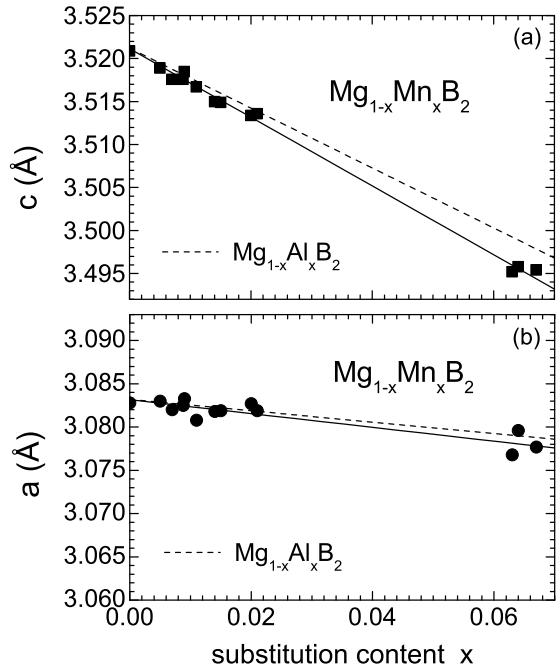}
\caption{Lattice parameters $c$ and $a$ versus Mn content $x$
(determined with EDX) for Mg$_{1-x}$Mn$_x$B$_2$ single crystals
(closed symbols, solid lines). The crystals are superconducting
for $x < 0.02$. The solid lines are linear fits to the data. The
dashed lines represent the lattice parameters for Al-substituted
crystals.\cite{Karpinski'PRB'2005}} \label{c,a(x)}
\end{figure}

In Fig.~\ref{c,a(x)} we show the lattice parameters $c$ and $a$
versus Mn content for Mg$_{1-x}$Mn$_x$B$_2$ single crystals with
$x$ from 0 to 0.067. A significant linear decrease of $4 \cdot
10^{-3}$~\AA/$\%$Mn of the $c$-axis parameter with substitution is
observed. Much weaker substitution effect on the $c$-axis
parameter was found for nearly single-phase polycrystalline
Mg$_{1-x}$Mn$_x$B$_2$, where the Mn content was taken as the
nominal content and thus could be
overestimated.\cite{Xu'JPSJ70'2001} The variation of the $a$-axis
parameter with $x$ is much smaller. Similar behavior of $c(x)$ and
$a(x)$ was reported for Al-substituted
crystals,\cite{Karpinski'PRB'2005} and for Co- and Cr-substituted
polycrystalline materials.\cite{Kuhberger'PhysicaC'2002,
Zhang'PhysicaC'2005} The distinct contraction of the MgB$_2$ unit
cell along the $c$-axis observed for our substituted crystals
indicates that Mn enters the crystal structure. Similar conclusion
has been also derived from the single crystal X-ray investigations
where it was possible to refine the Mg$_{1-x}$Mn$_x$B$_2$
structure with Mn on Mg position only. Considering the contraction
of the unit cell with Mn substitution, a simple comparison of the
ionic radii of Mg and Mn suggests that the effective valence state
of Mn can be 3+ with low-spin as well as high-spin configuration
or 2+ with low-spin configuration only. As we will show later, the
magnetic and X-ray absorption studies reveal that the Mn ions
substituted for Mg are divalent and in the low-spin configuration.

\begin{figure}[!htb]
\includegraphics*[width=0.45\textwidth]{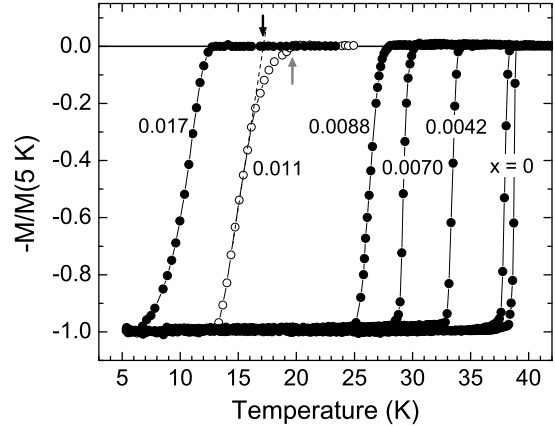}
\caption{Normalized magnetic moment $M$ versus temperature for the
Mg$_{1-x}$Mn$_x$B$_2$ single crystals with various Mn content $x$.
The measurements were performed in a field of 0.5~mT, after
cooling in a zero field. The superconducting transition
temperature $T_c$ is marked by a solid arrow and the transition
onset temperature $T_{co}$ by a grey arrow. Crystals with a sharp
transition (closed circles) were selected for further studies.}
\label{M(T)x}
\end{figure}

The superconducting transition temperature was determined from the
magnetic moment measurements performed as a function of
temperature in a 0.5~mT dc field in ZFC mode.  As an example, the
$M(T)$ results for crystals with various Mn content are shown in
Fig.~\ref{M(T)x}. The effective transition temperature $T_c$ and
the onset temperature $T_{co}$ were defined as illustrated in the
Figure. A broad transition to the superconducting state for the
crystal with $x =$ 0.011 is included to clearly illustrate the
definitions. A difference $\Delta T_c = T_{co} - T_c$ is
identified with the sample quality and it varies from 0.1 to 2.5~K
at 0.5~mT, depending on the Mn content and synthesis conditions.
Crystals with $\Delta T_c$(0.5~mT)~$\leq 1$~K were selected for
further studies.

Magnetic moment versus field has been measured to examine
shielding effects and to estimate an upper limit of the
superconducting volume fraction for Mn-substituted crystals.
Virgin magnetization curves $M(H)$ were obtained at low
temperatures for the Mg$_{1-x}$Mn$_x$B$_2$ crystals with $x =$
0.0088. For a crystal with a mass of 23.5~($\pm 0.5)$~$\mu$g and
dimensions $0.55 \times 0.35 \times 0.045$~mm$^3$, the
superconducting volume fraction $f =$ 0.96~($\pm 0.04$) was
derived at 10~K ($T < 0.5T_c$) with a demagnetizing factor $n =$
0.06 for $H$ parallel to the main surface of the crystal. This
confirms full diamagnetism of the Mn-substituted crystal
certifying its good quality. First deviation from the linear part
of the $M(H)$ virgin curve was used to roughly estimate the lower
critical field $\mu_oH_{c1} \simeq 19$ and 12~mT at 4.5 and 10~K,
respectively, for $H$ parallel to the $ab$-plane. These values
signify the upward curvature of the $H_{c1}$ versus $T$ dependence
(above 10~K) and are similar\cite{Zehetmayer'PRB'2002} or much
lower\cite{Perkins'SST'2002, Lyard'PRL'2004} than those observed
for unsubstituted MgB$_2$.

\begin{figure}[!htb]
\includegraphics*[width=0.45\textwidth]{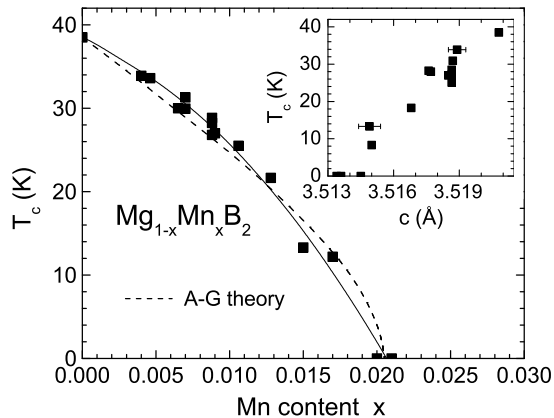}
\caption{Suppression of $T_c$ with Mn substitution for
Mg$_{1-x}$Mn$_x$B$_2$ single crystals with a sharp transition to
the superconducting state (see Fig.~\ref{M(T)x}). The solid line
is a polynomial fit to the experimental points. The dashed line
shows $T_c(x)$ predicted by the A-G pair-breaking theory. The
inset shows $T_c$ versus the lattice parameter $c$. For clarity,
only two error bars are shown.} \label{Tc(x,c)}
\end{figure}

The superconducting transition temperature systematically
decreases with Mn substitution resulting in a complete suppression
of superconductivity at $x \simeq$ 0.02, as shown in
Fig.~\ref{Tc(x,c)}. The suppression is faster than linear in the
whole range of doping and can be described by the magnetic
pair-breaking effect. According to the Abrikosov-Gor'kov (AG)
pair-breaking theory the interaction of magnetic impurities with
conduction electrons may break the time-reversal symmetry of the
Cooper-pairs and result in a rapid decrease of $T_c$ with the
concentration of magnetic ions $x$.\cite{Abrikosov'ZETF'1960} The
reduced $T_c(x)$ is well described by the relation $\ln(t_c) =
\Psi(\frac{1}{2}) - \Psi(\frac{1}{2}+0.14t_c\alpha/\alpha_{cr})$,
with $t_c = T_c(x)/T_c(0)$.\cite{Parks'Wallace'1969} $\Psi(z)$ is
digamma function, and $\alpha/\alpha_{cr}$ is the normalized
pair-breaking parameter which is identical to $x/x_{cr}$, where
$x_{cr}$ is the concentration of magnetic impurities required to
supperss $T_c$ to zero. The dependence of $T_c(x)/T_c(0)$ as a
function of $x/x_{cr}$ follows an universal relation. For small
$x$, $T_c(x)$ changes roughly linearly and drops more rapidly for
$x$ closer to $x_{cr}$. For the Mn-substituted MgB$_2$ crystals,
we found a very similar dependence (see Fig.~\ref{Tc(x,c)}),
however a small deviation from the A-G curve seems to be present.
This possible deviation could be a result of unconventional
two-gap superconductivity, where the interband scattering, which
may grow with the amount of substituted magnetic ions, is
postulated as an additional mechanism that reduces
$T_c$.\cite{Golubov'JPCM'2002, Mitrovic'JPCM'2004,
Kortus'PRB'2005, Dolgov'PRB'2005}

\begin{figure}[!htb]
\includegraphics*[width=0.45\textwidth]{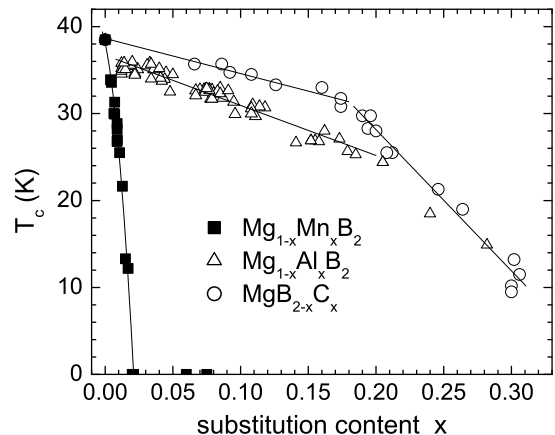}
\caption{Variation of $T_c$ for MgB$_2$ single crystals
substituted with nonmagnetic (Al, C) and magnetic (Mn) ions. The
most striking result is a rapid suppression of $T_c$ due to the
substitution of isovalent Mn for Mg. The main part of the results
on Al- and C-substituted crystals has been published in
Ref.~\onlinecite{Karpinski'PRB'2005} and
\onlinecite{Kazakov'PRB'2005}, respectively.} \label{Tc(x)}
\end{figure}

The rapid decrease of $T_c$ caused by Mn ions is particularly
clear when the Mn substitution is compared with others. In
Fig.~\ref{Tc(x)} we show $T_c(x)$ for MgB$_2$ crystals substituted
with Mn for Mg, and electron-doped Al for Mg and C for B. The
dramatic suppression of $T_c$ for the Mn-substituted crystals
seems to be a pure magnetic pair-braking effect, because any
essential changes of the electronic structure are not expected for
the reason that Mn substitutes as isovalent Mn$^{2+}$, as we
discuss below. This requires strong interaction between the
localized $3d$ electrons of Mn and conduction (mostly) $2s2p$
electrons of B even if we realize that the magnetic impurities are
located at the Mg sites, which are spatially separated from the B
planes. A large difference between the $T_c$ suppression rates for
magnetic and nonmagnetic substitutions is consistent with
orthogonality of the $\sigma$ and $\pi$ orbitals and,
consequently, with the much smaller interband than intraband
scattering.\cite{Mazin'PRL'2002} An interesting issue is if so
fine substitution of Mn for Mg, which yet changes $T_c$ so
rapidly, modifies the $\sigma$ and $\pi$ intraband scattering and
influences the $\sigma$-band anisotropy. This we discuss in the
paragraphs devoted to properties of the upper critical field.

\begin{figure}[!htb]
\includegraphics*[width=0.45\textwidth]{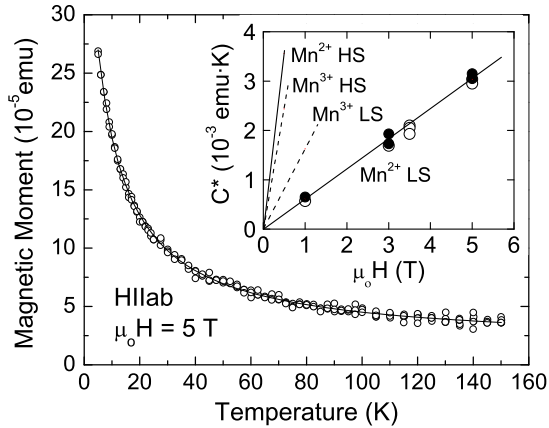}
\caption{Temperature dependence of the magnetic moment $M$ at a
constant field $\mu_oH =$ 5~T for MgB$_2$ single crystals
substituted with 0.88\% of Mn. The sample consists of 25 crystals
with $T_c \simeq 28 (\pm 1)$~K and a total mass of 847~$\mu$g
($1.61\cdot10^{-7}$ mol Mn). The inset shows the Curie part
$C^\star$ of the $M(T)$ dependence as obtained from the experiment
(circles). The lines are the expectations for $C^\star$ based on
Mn$^{2+}$ (solid lines) or Mn$^{3+}$ (dashed lines) in the
high-spin (HS) and low-spin (LS) configuration. The measurements
reveal Mn to be divalent in the low-spin configuration. $H$ was
oriented either parallel to the $ab$-plane (open circles) or
70$^\circ$ off the $ab$-plane (solid circles).} \label{C(H)}
\end{figure}

The normal state magnetization was measured on individual single
crystals (typically $m \simeq 25~\mu$g) and on an assembly of 25
crystals (0.88$\%$~Mn, total $m \simeq 847~\mu$g) attached with
vacuum grease to a nonmagnetic sample holder. The presence of Mn
ions manifests itself in a Curie-Weiss contribution that dominates
$M(T)$ even at the low Mn concentrations of $< 1\%$. An example of
the $M(T)$ dependence is shown in Fig.~\ref{C(H)} for the
multi-crystal assembly. For individual crystals, $M(H)$ curves
were measured at various temperatures to calculate $M(T)$, since
the small crystal mass and the low Mn content resulted in $\sim
10^{-8}$ - $10^{-7}$ mol Mn. The normal state magnetic moment was
analyzed according to the formula $M(T) = M_o +
C^\star/(T+\Theta)$, where $C^\star/(T+\Theta)$ is the Curie-Weiss
contribution associated with the Mn local moments. The effective
interaction temperature $\Theta$ is found to be $\leq 2$~K,
reflecting the high dilution of the Mn ions. The value of
$C^\star$ is shown in the inset of Figure~\ref{C(H)} for $M(T)$
measurements in fields up to 5~T, and for $H$ either parallel to
or 70$^\circ$ off the crystal $ab-$plane. $C^\star(H)$ is
isotropic within the measurement limits and grows linearly with
$H$, as expected.

The magnitude of the local-moment Curie-Weiss part of $M(T)$
reveals unambiguously that Mn in MgB$_2$ is in the divalent state,
isovalent to Mg. The lines in the inset of Fig.~\ref{C(H)} are the
calculated values of $C^\star(H)$, assuming Mn$^{2+}$ and
Mn$^{3+}$, in either high-spin or low-spin state. The measured
data, which correspond to $m_m \simeq$ 1.7~$\mu_B$ per Mn ion, are
in excellent agreement with low-spin Mn$^{2+}$, and they are
clearly distinct from the alternative states. Thus, we conclude
that the crystal field acting upon the $d$ electrons is strong
enough to produce a low-spin configuration with an effective $S =
1/2$. Measurements on other crystals, including one with
6.5$\%$~Mn, lead to the same conclusion.

\begin{figure}[!htb]
\includegraphics*[width=0.45\textwidth]{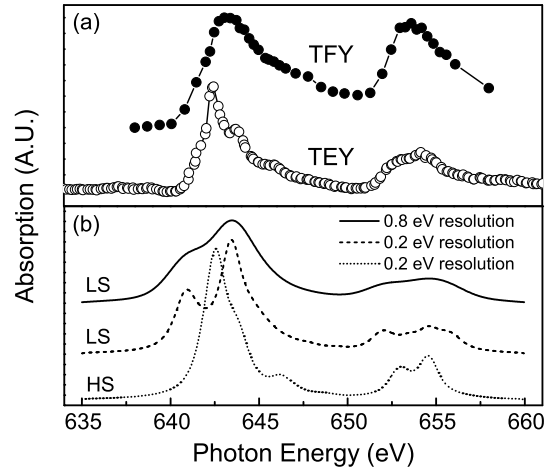}
\caption{(a) XAS spectra for the TEY and the TFY experiments on
MgB$_2$ single crystals substituted with 6.7\% of Mn. In the TEY
mode, the spectrum is probably dominated by a MnO surface layer,
while the TFY mode is more sensitive to the bulk. (b) Atomic model
calculations for the $3d^5$ ground state of Mn in the low-spin
(LS) and high-spin (HS) configuration. All the curves are shifted
for clarity.} \label{Abs(ph-en)}
\end{figure}

Information on the $d$-electron configuration can be deduced also
from X-ray absorption spectroscopy (XAS). The $2p$ to $3d$ XAS
spectrum in transition metals has been proved to be a sensitive
probe to their electronic ground state.\cite{Thole'PRB'1985} It is
possible to calculate the XAS spectrum with a standard Cowan
code,\cite{Cowan'UC'1981} based on an atomic model, and compare it
to the experimental spectra. This approach is very suitable for
the determination of the valency of transition metal impurities in
crystals. In Fig.~\ref{Abs(ph-en)} we show the experimental TEY
and TFY spectra (see Experimental) together with the atomic model
calculations without and with a cubic crystal field, which mimics
the presence of the solid around the Mn ion. In the upper panel of
the figure the TEY spectrum shows the typical shape of the
high-spin Hund's rule ground state. The TFY measurements, though
the resolution doesn't allow to distinguish very clear features,
shows a shoulder on the low energy side of the $L_2$ edge and a
shift of the white line by about 2~eV towards higher energies. In
the lower panel we plot three simulations of the Mn$^{2+}$ XAS
spectrum without considering any crystal field and with a cubic
crystal field just above the high-spin to low-spin transition
value, which for Mn is around 2.4~eV. The ligand field value at
the Mn site has been obtained performing a band calculation
assuming the MgB$_2$ crystal structure and replacing all the Mg
atoms with Mn atoms; this calculation gives a crystal field value
of about 2~eV, which as a first approximation is close enough to
the value necessary to induce the high-spin to low-spin transition
in Mn. The effect of the crystal field on the spectrum is to shift
the white line at higher energies by an amount which is related to
the crystal field value itself; a peak on the low energy side of
the spectrum also arises, which in a spectrum with lower
resolution becomes a pronounced shoulder. Our interpretation is
that a layer of MnO is probably present on the surface of the
sample, giving to the TEY spectrum the typical shape of the
high-spin ground state. However, the TFY spectrum reveals the bulk
properties of the sample, suggesting that the Mn$^{2+}$ ions are
in a low-spin configuration induced by the crystal field effect.
This result is consistent with our magnetic measurements,
discussed above.

\begin{figure}[!htb]
\includegraphics*[width=0.45\textwidth]{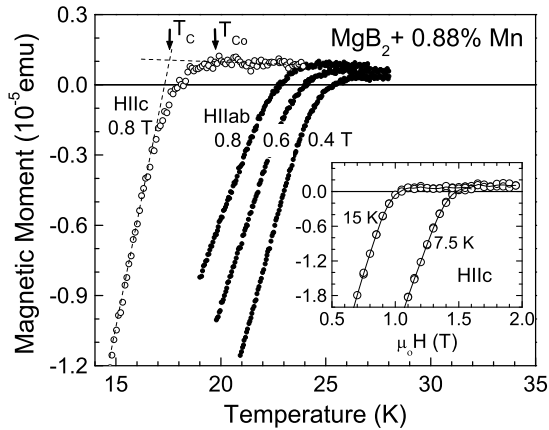}
\caption{Diamagnetic response of the MgB$_2$ crystal substituted
with 0.88\% of Mn. Shown are examples of measurements in constant
field or at constant temperature (inset), for the two main
orientations of the field. The transition temperature $T_c$ and
the transition onset temperature $T_{co}$ are marked by arrows.}
\label{M(T,H)}
\end{figure}

\begin{figure}[!htb]
\includegraphics*[width=0.45\textwidth]{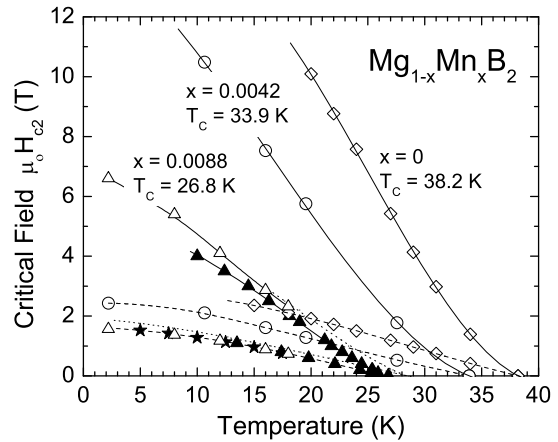}
\caption{Upper critical field $H_{c2}$ versus temperature for the
Mg$_{1-x}$Mn$_x$B$_2$ single crystals with $x =$ 0 (diamonds),
0.0042 (circles) and 0.0088 (triangles, stars). The $H_{c2}(T)$
data were obtained by magnetization measurements (solid symbols)
or derived from torque measurements (open symbols), for the
magnetic field $H$ oriented parallel (solid lines) and
perpendicular (dashed lines) to the $ab$-plane. The magnetization
measurements were performed at constant $H$ with increasing $T$
(solid triangles) or at constant $T$ with increasing $H$ (stars).
The doted lines show the $H_{c2}(T)$ corresponding to the
transition onset temperature $T_{co}$ (see Fig.~\ref{M(T,H)}). The
lines are a guide for the eye.} \label{Hc2(T)}
\end{figure}

The upper critical field, $H_{c2}$, has been determined from
magnetic moment measurements performed as a function of
temperature at constant field or versus field at constant
temperature. In Fig.~\ref{M(T,H)} we show examples of $M(T)$ and
$M(H)$ results for the crystal substituted with 0.88\% of Mn. The
results have been obtained with a field oriented parallel,
$H^{ab}$, and perpendicular, $H^{c}$, to the $ab$-plane. The
superconducting transition temperature $T_c$ and the transition
onset temperature $T_{co}$ have been defined as shown in the
Figure. The difference between $T_c$ and $T_{co}$ obviously
depends on the field orientation and value, however this modifies
the $H_{c2}(T)$ results only slightly (see Fig.~\ref{Hc2(T)}).
Extensive sets of data similar to these presented in
Fig.~\ref{M(T,H)} are analyzed to construct the $H_{c2}$-$T$ phase
diagram. Figure~\ref{Hc2(T)} shows the upper critical field of the
Mg$_{1-x}$Mn$_x$B$_2$ crystals with $x =$ 0.0042 and 0.0088, and,
for comparison, of the unsubstituted compound. For the heavily
doped crystal, special attention has been paid to obtain accurate
$H_{c2}$ values at low fields to determine the upper critical
field slope, d$H_{c2}$/d$T$, near $T_c$. For this crystal ($x =$
0.0088, $T_c =$ 26.8~K), d$\mu_oH_{c2}$/d$T$ at $T_c$ is equal to
-0.205($\pm0.005$) and -0.100($\pm0.005$)~T/K, for $H$ oriented
parallel and perpendicular to the $ab$-plane, respectively. These
values are practically the same as for unsubstituted crystals:
-0.21 and -0.10~T/K, respectively.

According to the quasi-classic model of the two-band
superconductor in the dirty limit (without magnetic impurities),
the temperature dependence of $H_{c2}$ close $T_c$ is determined
by the intraband scattering for the band with a maximum
diffusivity, when assumed that the intraband and interband
electron-phonon coupling constants are finite.\cite{Suhl'PRL'1959,
Gurevich'PRB'2003} For pure MgB$_2$, the band with a maximum
diffusivity is the $\pi$-band.\cite{Mazin'PRL'2002,
Putti-Braccini'PRB'2003, Putti'SST'2003} Thus, the unchanged
d$H_{c2}$/d$T$, observed for the Mn-substituted MgB$_2$ crystals
close to $T_c$, suggests that the $\pi$-band diffusivity (or
scattering) is not affected by the low-level substitution of
magnetic isovalent ions for Mg. As a consequence, the diffusivity
in the $\pi$-band remains dominant and the upper critical field at
zero temperature, $H_{c2}(0)$, should be determined by $T_c$ and
the minimum diffusivity,\cite{Gurevich'PRB'2003} i.e., the
diffusivity in the $\sigma$-band. A roughly linear
$H_{c2}(0)$-$T_c$ relation is observed for the Mn-substituted
crystals. Thus, the Mn substitution causes merely minimal changes
in the $\pi$-band and the $\sigma$-band diffusivity, and
suppresses $T_c$ by spin-flip scattering. Similar conclusions have
been drown from a point-contact
spectroscopy.\cite{Gonnelli'cond-mat'0510329}

There are at least two mechanisms possible for the reduction of
$T_c$ in MgB$_2$ substituted with magnetic isovalent ions. One is
the impurity-induced (nonmagnetic) interband scattering and the
second is the magnetic pair-breaking effect. The interband
scattering alone is expected to reduce $T_c$ at most to about 25~K
and, most likely, for the amount of substituted ions much higher
than 2\%,\cite{Mitrovic'JPCM'2004, Kortus'PRB'2005} that for the
Mn-substituted crystals suppresses $T_c$ to zero. Moreover, any
significant modification of the interband scattering requires a
substantial concentration of the impurity ions in the B plane
rather than in the Mg plane,\cite{Ervin'PRB'2003, Mazin'PRL'2002}
as shown for C-substituted MgB$_2$.\cite{Kortus'PRB'2005,
Gonnelli'PRB'2005} Thus, the main mechanism that controls $T_c$ in
the Mn-substituted MgB$_2$ remains the magnetic pair-braking
effect.

\begin{figure}[!htb]
\includegraphics*[width=0.45\textwidth]{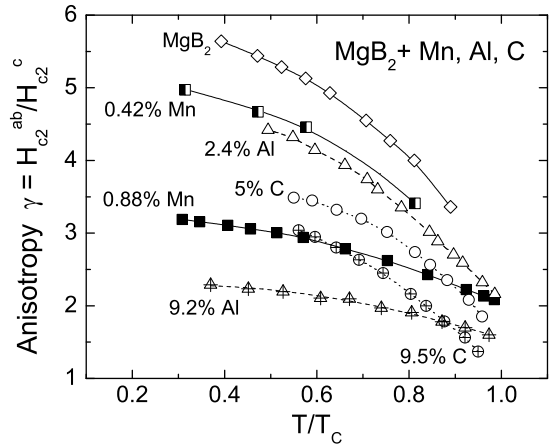}
\caption{Upper critical field anisotropy versus reduced
temperature for the MgB$_2$ unsubstituted single crystals
(diamonds; $T_c =$ 38.2~K) and substituted with Mn (squares;
0.42\% Mn, $T_c =$ 33.9~K; 0.88\% Mn, $T_c =$ 26.8~K), Al
(triangles; 2.4\% Al, 35.3~K; 9.2\% Al, 32.0~K) and C (circles;
5\% C, 34.3~K; 9.5\% C, 30.1~K). The data for Al- and
C-substituted crystals were derived from $H_{c2}(T)$ results
published in Ref.~\onlinecite{Karpinski'PRB'2005} and
\onlinecite{Kazakov'PRB'2005}.} \label{gamma(t)}
\end{figure}

The overall temperature dependence of $H_{c2}$ is characterized
mainly by a reduction of the scales when Mn is substituted. In
particular, the anisotropy remains well pronounced, and so does
the marked up-turn of $H_{c2}^{ab}$ below $T_c$, which means that
the two-band character determining $H_{c2}(T)$ is essentially
unaffected by Mn substitution. This is in line with the
above-noted unchanged initial slope of d$H_{c2}$/d$T$. In
Fig.~\ref{gamma(t)} we show the temperature dependence of the
upper critical field anisotropy, $\gamma = H_{c2}^{ab}/H_{c2}^c$,
for the MgB$_2$ single crystals substituted with 0.42\% and 0.88\%
of Mn. For comparison, the anisotropy for non-substituted,
Al-substituted, and C-substituted crystals is also presented. At
low temperatures, a large reduction of $\gamma$ from 6 to 3.3 is
observed for the crystal with 0.88\% of Mn. Along with the
lowering of $\gamma$, its temperature dependence weakens. This
behavior, observed for the MgB$_2$ crystals substituted with
magnetic isovalent Mn$^{2+}$, is similar to that obtained for the
crystals substituted with electron-adding Al$^{3+}$. Similar
$\gamma(T)$ dependencies are observed for crystals with similar
$T_c$'s but with much different Mn and Al contents (lower
substitution) or for crystals with significantly different $T_c$'s
(heavier substitution, see Fig.~\ref{gamma(t)}). For example,
0.42\% ($\sim$1\%) of substituted Mn results in changes similar to
those observed for 2.4\% ($\sim~$9\%) of substituted Al. Thus, the
mechanism that is responsible for the reduction of the anisotropy
and for changes of its temperature dependence has to be different
in the both cases. Note, that the temperature dependence of
$\gamma$ obtained for the Mn- and Al-substituted crystals differs
significantly from that derived for the C-substituted crystals.
This we discuss shortly in the next paragraph.

The upper critical field anisotropy decreases with increasing
temperature for both unsubstituted and substituted crystals, as
shown in Fig.~\ref{gamma(t)}. For a weak-coupling multiband BCS
model for two-gap superconductors (without magnetic impurities), a
negative anisotropy slope, d$\gamma$/d$T$, is expected for the
case when diffusivity in the $\pi$-band
dominates.\cite{Gurevich'PRB'2003} This requirement seems to be
fulfilled in the clean non-substituted or C-substituted MgB$_2$,
where C on the B position decreases the diffusivity mainly in the
$\sigma$-band, as shown for C-substituted single
crystals\cite{Sologubenko'PRB'2005} and epitaxial thin
films.\cite{Pallecchi'cond-mat'0508509} Thus, the negative slope
d$\gamma$/d$t$, which at lower temperatures ($t = T/T_c \leq 0.5$)
is similar for non-substituted and C-substituted crystals (see
Fig.~\ref{gamma(t)}), is fully consistent with this prediction. On
the other hand, when the diffusivity in the $\sigma$-band
dominates, $\gamma$($T$) is expected to be less temperature
dependent, or d$\gamma$/d$t$ may even become
positive.\cite{Gurevich'PRB'2003,Kogan'PRB'2004} The diffusivity
in the $\sigma$-band may dominate, when the scattering in the
$\pi$-band increases substantially, e.g., due to the substitution
of Al for Mg.\cite{Carrington'cond-mat'0504664} Both Mn and Al
substitutions show a tendency to lower $\mid$d$\gamma$/d$t$$\mid$
with the increasing amount of substituted ions. For an
unsubstituted crystal ($T_c =$ 38.2~K), $\mid$d$\gamma$/d$t$$\mid
=$ 2.6 at $t = 0.4$ and decreases slightly to about 2.3, for the
crystals with 0.42\% of Mn ($T_c =$ 33.9~K) or 2.4\% of Al ($T_c
=$ 35.3~K), and more significantly to 0.91 and 0.42, for the
crystals with 0.88\% of Mn ($T_c =$ 26.8~K) and 9.2\% of Al ($T_c
=$ 32.0~K), respectively.

For the Al-substituted crystals, the observed suppression of
$\mid$d$\gamma$/d$t$$\mid$ can be interpreted as a result of
increased intraband scattering in the
$\pi$-band.\cite{Carrington'cond-mat'0504664} For the
Mn-substituted crystals, the explanation seems to be different, as
d$H_{c2}$/d$T$ at $T_c$ remains unchanged. Here, pair breaking due
to spin-flip scattering appears to dominate the reduction of $T_c$
and the $H_{c2}$ anisotropy. A reduction of the energy gap due to
spin-flip scattering is expected to be different for both bands
and thus the ratio $\Delta_\sigma/\Delta_\pi$ may vary with
temperature in a way that is different from that in unsubstituted
MgB$_2$. The details can be worked out through $T$-dependent gap
spectroscopy revealing changes in the magnitude and weakening of
the temperature dependence of $\Delta_\sigma$ and $\Delta_\pi$ due
to spin-flip scattering. Such spectroscopic studies may also
reveal if some of the scenarios discussed theoretically for
magnetic pair breaking in MgB$_2$ apply to this
compound.\cite{Moca'PRB'2002}

Studies of stronger Mn- and Al-substituted single crystals are in
progress to examine, among others, the question if d$\gamma$/d$T$
changes to positive, when the $\sigma$-band diffusivity becomes
larger than the $\pi$-band diffusivity, or saturates at the value
$\simeq 0$, as expected for a single-gap superconductor. Recently,
d$\gamma$/d$T >$ 0 has been observed for Al-substituted
polycrystalline Mg$_{1-x}$Al$_x$B$_2$ with $x =$
0.2.\cite{Angst'PRB71'2005} This suggests that for larger $x$, an
inversion of the two-band hierarchy may appear, as a result of
changes in the interband pairing strength, due to appropriate
modifications of the Coulomb
pseudopotential.\cite{Bussmann-Holder'PRB'2003,
Ummarino'PhysicaC'2004} The inversion of the two-band hierarchy
has just been observed by electron energy-loss spectroscopy for
heavily substituted polycrystalline Mg$_{1-x}$Al$_x$B$_2$ with $x
>$ 0.33.\cite{Cooley'cond-mat'0504463} For the C-substituted
crystals, an almost parallel shift of the $\gamma(T)$ dependence
is observed. Thus, it will be interesting to see if $\gamma$ drops
below 1 ($H_{c2}^{ab} < H_{c2}^c$), for crystals with C content
larger than 10\%, when the intraband and interband electron-phonon
coupling constants are predicted to decrease and become
similar.\cite{Ummarino'PRB'2005}

%---------------------------------------------------------------------
\section{CONCLUSIONS}

We have studied the influence of Mn substitution on the
superconducting properties of MgB$_2$ by growing single crystals
with Mn concentrations up to 7\%, and measuring their magnetic
properties and X-ray absorption spectra. Mn suppresses $T_c$ very
effectively at an initial rate of $\sim 10$~K/\%Mn, and $T_c$ is
fully suppressed at $\simeq 2$\% of Mn. The temperature dependence
of $H_{c2}(T)$ and $\gamma(T)$ obtained for the Mn-substituted
single crystals is similar to that reported previously for MgB$_2$
substituted with nonmagnetic Al, provided that the crystals with
similar $T_c$ are compared. This suggests that in MgB$_2$, where
Mg is substituted with magnetic or nonmagnetic ions, the main
parameter that controls both $H_{c2}(T)$ and $\gamma(T)$ is the
superconducting transition temperature, irrespective of the
mechanism responsible for the $T_c$ suppression. For
Mn-substituted MgB$_2$, this suppression is found to be due to the
magnetic pair-breaking effect caused by Mn ions, as Mn substitutes
for Mg isovalently as Mn$^{2+}$ in the low-spin ($S =$ 1/2)
configuration. Along with the reduction of $T_c$, the upper
critical field $H_{c2}(0)$ and its anisotropy are also reduced,
while the initial slope d$H_{c2}$/d$T$ near $T_c$ and the
associated anisotropy remain essentially unaffected. These results
suggest that the magnetic Mn substitution predominantly influences
the superconducting properties through spin-flip scattering,
leaving the diffusivity in the $\sigma$ and $\pi$ bands largely
unaffected. A further treatment will have to include the detailed
knowledge of the influence of magnetic pair breaking on the
$\sigma$ and $\pi$ bands in MgB$_2$ and the resulting modification
of the electronic properties of this two-gap superconductor.

%---------------------------------------------------------------------
\begin{acknowledgments}

This work was supported in part by the Swiss National Science
Foundation, MaNEP, the Centre of Excellence (CELTAM) established
under the grant No. GMA1-2002-72017 within 5$^{th}$ Framework
Program of European Community (contract No. G5MA-CT.2002.04032)
(KR), and the Polish State Committee for Scientific Research under
a research project for the years 2004-2006 (P03B 037 27) (RP and
AW). The SMART CCD measurements were performed at the Laboratory
of Inorganic Chemistry, ETH Zurich.

\end{acknowledgments}

%---------------------------------------------------------------------


\begin{thebibliography}{99}

\bibitem{Maple'AP'1976} For a review, see M. B. Maple, Appl. Phys.
\textbf{9}, 179 (1976).

\bibitem{Maple'Fischer'1982} For a review, see \emph{
Superconductivity in Ternary Compounds II, Topics in Current
Physics}, edited by M. B. Maple and \O. Fischer (Springer-Verlag,
Berlin, Heidelberg, New York 1982), Vol. 34.

\bibitem{Canfield'PT'1998} For a review, see P. C. Canfield,
P. L. Gammel, and D. J. Bishop, Phys. Today \textbf{51} 40 (1998).

\bibitem{Muller'RPrPh'2001} For a review, see K. H. M\"{u}ller and
V. N. Narozhnyi, Rep. Prog. Phys. \textbf{64} 943 (2001).

\bibitem{Flouquet'cond-mat'0505713} J. Flouquet, G. Knebel,
D. Braithwaite, D. Aoki, J. P. Brison, F. Hardy, A. Huxley, S.
Raymond, B. Salce, and I. Sheikin, cond-mat/0505713 (unpublished),
and references cited therein.

\bibitem{Maple'PhysicaB+C'1987} M. B. Maple, Y. Dalichaouch,
J. M. Ferreira, R. R. Hake, B. W. Lee, J. J. Neumeier, M. S.
Torikachvili, K. N. Yang, H. Zhou, R. P. Guertin, and M. V. Kuric,
Physica \textbf{148B}, 155 (1987).

\bibitem{Maple'JLCM'1989} M. B. Maple, J. M. Ferreira,
R. R. Hake, B. W. Lee, J. J. Neumeier, C. L. Seaman, K. N. Yang,
and H. Zhou, J. Less-Common Metals \textbf{149}, 405 (1989).

\bibitem{Nachtrab'cond-mat'0508044} For a recent review, see
T. Nachtrab, C. Bernhard, C. T. Lin, D. Koelle, and R. Kleiner,
cond-mat/0508044 (unpublished).

\bibitem{Ishida'Nature'1998} K. Ishida, H. Mukuda, Y. Kitaoka,
K. Asayama, Z. Q. Mao, Y. Mori, and Y. Maeno, Nature \textbf{396},
658 (1998).

\bibitem{Saxena'Nature'2000} S. S. Saxena, P. Agarwal, K. Ahilan,
F. M. Grosche, R. K. W. Haselwimmer, M. J. Steiner, E. Pugh, I. R.
Walker, S. R. Julian, P. Monthoux, G. G. Lonzarich, A. Huxley, I.
Sheikin, D. Braithwaite, and J. Flouquet, Nature \textbf{406}, 587
(2000).

\bibitem{Pfleiderer'Nature'2001} C. Pfleiderer, M. Uhlarz,
S. M. Hayden, R. Vollmer, H. v. L\"{o}hneysen, N. R. Bernhoeft,
and G. G. Lonzarich, Nature \textbf{412}, 58 (2001).

\bibitem{Nagamatsu'Nature'2001} J. Nagamatsu, N. Nakagawa,
T. Muranaka, Y. Zenitani, and J. Akimitsu, Nature \textbf{410}, 63
(2001).

\bibitem{Kortus'PRL86'2001} J. Kortus, I. I. Mazin,
K. D. Belashchenko, V. P. Antropov, and L. L. Boyer, Phys. Rev.
Lett. \textbf{86}, 4656 (2001).

\bibitem{Liu'PRL'2001} A. Y. Liu, I. I. Mazin, and J. Kortus,
Phys. Rev. Lett. \textbf{87}, 087005 (2001).

\bibitem{Choi'Nature'2002} H. J. Choi, D. Roundy, H. Sun,
M. L. Cohen, and S. G. Louie, Nature \textbf{418}, 758 (2002).

\bibitem{Canfield'PT'2003} For a review, see P. C. Canfield and
G. W. Crabtree, Phys. Today \textbf{56}, 34 (2003).

\bibitem{Pena'PRB'2002} O. de la Pe\~{n}a, A. Aguayo, and R. de
Coss, Phys. Rev. B \textbf{66}, 012511 (2002).

\bibitem{Kasinathan'PhysicaC'2005} D. Kasinathan, K.-W. Lee, and
W. E. Pickett, Physica C \textbf{424}, 116 (2005).

\bibitem{Klie'cond-mat'0510002} R. F. Klie, J. C. Zheng, Y. Zhu,
A. J. Zambano, and L. D. Cooley, cond-mat/0510002 (unpublished).

\bibitem{Golubov'JPCM'2002} A. A. Golubov, J. Kortus, O. V. Dolgov,
O. Jepsen, Y. Kong, O. K. Andersen, B. J. Gibson, K. Ahn, and R.
K. Kremer, J. Phys.: Condens. Matter \textbf{14}, 1353 (2002).

\bibitem{Mitrovic'JPCM'2004} B. Mitrovi\'{c}, J. Phys.: Condens.
Matter \textbf{16}, 9013 (2004).

\bibitem{Kortus'PRB'2005} J. Kortus, O. V. Dolgov, R. K. Kremer,
and A. A. Golubov, Phys. Rev. Lett. \textbf{94}, 027002 (2005).

\bibitem{Dolgov'PRB'2005} O. V. Dolgov, R. K. Kremer, J. Kortus,
A. A. Golubov, and S. V. Shulga, Phys. Rev. B \textbf{72}, 024504
(2005).

\bibitem{Golubov'PRB'1997} A. A. Golubov and I. I. Mazin,
Phys. Rev. B \textbf{55}, 15146 (1997).

\bibitem{Angst'PRB71'2005} M. Angst, S. L. Bud'ko, R. H. T. Wilke,
and P. C. Canfield, Phys. Rev. B \textbf{71}, 144512 (2005).

\bibitem{Moca'PRB'2002} C. P. Moca and C. Horea, Phys. Rev.B
\textbf{66}, 052501 (2002).

\bibitem{Fischer'ApplPhys'1978} For a review, see \O. Fischer,
Appl. Phys. \textbf{16}, 1 (1978).

\bibitem{Shrivastava'PhysRep'1984} For a review, see K. N.
Shrivastava and K. P. Sinha, Phys. Reports \textbf{115}, 93
(1984).

\bibitem{Goldman'PRB'1994} A. I. Goldman, C. Stassis, P. C. Canfield,
J. Zarestky, P. Dervenagas, B. K. Cho, and D. C. Johnston, Phys.
Rev. B \textbf{50}, 9668 (1994).

\bibitem{Chervenak'PRB'1995} J. A. Chervenak and J. M. Valles,
Phys. Rev. B \textbf{51}, 11977 (1995).

\bibitem{Bill'PhysicaC'1998} A. Bill, S. A. Wolf, Yu. N.
Ovchinnikov, and V. Z. Kresin, Physica C \textbf{298}, 231 (1998).

\bibitem{Ghosh'PRB'2001} G. Ghosh, A. D. Chinchure, R. Nagarajan,
C. Godart, and L. C. Gupta, Phys. Rev. B \textbf{63}, 212505
(2001).

\bibitem{Kim'PRB'2002} Chang-An Kim and B. K. Cho, Phys. Rev. B
\textbf{66}, 214501 (2002).

\bibitem{Xiao'PRB'1990} G. Xiao, M. Z. Cieplak, J. Q. Xiao, and C.
L. Chien, Phys. Rev. B \textbf{42}, 8752 (1990).

\bibitem{Szabo'PRB'2000} P. Szab\'{o}, P. Samuely, A. G. M.
Jansen, J. Marcus, and P. Wyder, Phys. Rev. B \textbf{62}, 3502
(2000).

\bibitem{Chattopadhyay'JPCM'2002} A. K. Chattopadhyay, R. A.
Klemm, and D. Sa, J. Phys.: Condens. Matter \textbf{14}, L577
(2002).

\bibitem{Poddar'EPJB'2003} A. Poddar and B. Chattopadhyay, Eur.
Phys. J. B \textbf{35}, 69 (2003).

\bibitem{Rogacki'PRB'2003} K. Rogacki, Phys. Rev. B \textbf{68},
100507 (2003).

\bibitem{Slusky'Nature'2001} J. S. Slusky, N. Rogado, K. A. Regan,
M. A. Hayward, P. Khalifah, T. He, K. Inumaru, S. M. Loureiro, M.
K. Haas, H. W. Zandbergen, and R. J. Cava, Nature \textbf{410},
343 (2001).

\bibitem{Cava'PhysicaC'2003} R. J. Cava, H. W. Zandbergen, and K.
Inumaru, Physica C \textbf{385}, 8 (2003).

\bibitem{Putti'PRB'2003} M. Putti, M. Affronte, P. Manfrinetti,
and A. Palenzona, Phys. Rev. B \textbf{68}, 094514 (2003).

\bibitem{Karpinski'PRB'2005} J. Karpinski, N. D. Zhigadlo, G. Schuck,
S. M. Kazakov, B. Batlogg, K. Rogacki, R. Puzniak, J. Jun, E.
M\"{u}ller, P. W\"{a}gli, R. Gonnelli, D. Daghero, G. A. Ummarino,
and V. A. Stepanov, Phys. Rev. B \textbf{71}, 174506 (2005).

\bibitem{Zambano'SST'2005} A. J. Zambano, A. R. Moodenbaugh,
and L. D. Cooley, Supercond. Sci. Technol. \textbf{18}, 1411
(2005).

\bibitem{Bharathi'PhysicaC'2002} A. Bharathi, S. J. Balaselvi,
S. Kalavathi, G. L. N. Reddy, V. S. Sastry, Y. Hariharan, and T.
S. Radhakrishnan, Physica C \textbf{370}, 211 (2002).

\bibitem{Lee'PhysicaC'2003} S. Lee, T. Masui, A. Yamamoto,
H. Uchiyama, and S. Tajima, Physica C \textbf{397}, 7 (2003).

\bibitem{Masui'PRB'2004} T. Masui, S. Lee, and S. Tajima,
Phys. Rev. B \textbf{70}, 024504 (2004).

\bibitem{Kazakov'PRB'2005} S. M. Kazakov, R. Puzniak, K. Rogacki,
A. V. Mironov, N. D. Zhigadlo, J. Jun, Ch. Soltmann, B. Batlogg,
and J. Karpinski, Phys. Rev. B \textbf{71}, 024533 (2005).

\bibitem{Xu'JPSJ70'2001} S. Xu, Y. Moritomo, K. Kato, and A. Nakamura,
J. Phys. Soc. Japan \textbf{70}, 1889 (2001).

\bibitem{Kuhberger'PhysicaC'2002} M. K\"{u}hberger and G. Gritzner,
Physica C \textbf{370}, 39 (2002).

\bibitem{Dou'SST'2005} S. X. Dou, S. Soltanian, Y. Zhao, E. Getin,
Z. Chen, O. Shcherbakova, and J. Horvat, Supercond. Sci. Technol.
\textbf{18}, 710 (2005).

\bibitem{Zhang'PhysicaC'2005} H. Zhang, J. Zhao, and L. Shi,
Physica C \textbf{424}, 79 (2005).

\bibitem{Gonnelli'cond-mat'0510329} R. S. Gonnelli, D. Daghero,
G. A. Ummarino, A. Calzolari, M. Tortello, V. A. Stepanov, N. D.
Zhigadlo, K. Rogacki, and J. Karpinski, cond-mat/0510329
(unpublished).

\bibitem{Schuck'2006} For details, see G. Schuck et al., in preparation.

\bibitem{Angst'PRL'2002} M. Angst, R. Puzniak, A. Wisniewski, J. Jun,
S. M. Kazakov, J. Karpinski, J. Roos, and H. Keller, Phys. Rev.
Lett. \textbf{88}, 167004 (2002).

\bibitem{Zangrando'RSI'2001} M. Zangrando, M. Finazzi, G. Paolucci,
G. Comelli, B. Diviacco, R. P. Walker, D. Cocco, and F.
Parmigiani, Rev. Sci. Instrum. \textbf{72}, 1313 (2001).

\bibitem{Groot'SSC'1994} F. M. F. de Groot, M. A. Arrio, Ph. Sainctavit,
Ch. Cartier, and C. T. Chen, Solid State Commun. \textbf{92}, 991
(1994).

\bibitem{Zehetmayer'PRB'2002} M. Zehetmayer, M. Eisterer, J. Jun,
S. M. Kazakov, J. Karpinski, A. Wisniewski, and H. W. Weber, Phys.
Rev. B \textbf{66}, 052505 (2002).

\bibitem{Perkins'SST'2002} G. K. Perkins, J. Moore, Y. Bugoslavsky,
L. F. Cohen, J. Jun, S. M. Kazakov, J. Karpinski, and A. D.
Caplin, Supercond. Sci. Technol. \textbf{15}, 1156 (2002).

\bibitem{Lyard'PRL'2004} L. Lyard, P. Szab\'{o}, T. Klein, J. Marcus,
C. Marcenat, K. H. Kim, B. W. Kang, H. S. Lee, and S. I. Lee,
Phys. Rev. Lett. \textbf{92}, 057001 (2004).

\bibitem{Abrikosov'ZETF'1960} A. A. Abrikosov and L. P. Gor'kov,
ZETF \textbf{39}, 1781 (1960) [Sov. Phys. JETP. \textbf{12}, 1243
(1961)].

\bibitem{Parks'Wallace'1969} R. D. Parks, in \emph{Superconductivity},
edited by P. R. Wallace (Gordon and Breach, Science Publishers,
New York 1969), Vol. 2, P. 625.

\bibitem{Mazin'PRL'2002} I. I. Mazin, O. K. Andersen, O. Jepsen,
O. V. Dolgov, J. Kortus, A. A. Golubov, A. B. Kuz'menko, and D.
van der Marel, Phys. Rev. Lett. \textbf{89}, 107002 (2002).

\bibitem{Thole'PRB'1985} B. T. Thole, R. D. Cowan, G. A. Sawatzky,
J. Fink, and J. C. Fuggle, Phys. Rev. B, \textbf{31}, 6856 (1985).

\bibitem{Cowan'UC'1981} R. D. Cowan, \emph{The theory of Atomic
Structure and Spectra} (University of California press, Berkeley
1981).

\bibitem{Suhl'PRL'1959} H. Suhl, B. T. Matthias, and L. R. Walker,
Phys. Rev. Lett. \textbf{3}, 552 (1959).

\bibitem{Gurevich'PRB'2003} A. Gurevich, Phys. Rev. B \textbf{67},
184515 (2003).

\bibitem{Putti-Braccini'PRB'2003} M. Putti, V. Braccini,
E. Galleani d'Agliano, F. Napoli, I. Pallecchi, A. S. Siri, P.
Manfrinetti, and A. Palenzona, Phys. Rev. B \textbf{67}, 064505
(2003).

\bibitem{Putti'SST'2003} M. Putti, V. Braccini, E. Galleani,
F. Napoli, I. Pallecchi, A. S. Siri, P. Manfrinetti, and A.
Palenzona, Supercond. Sci. Technol. \textbf{16}, 188 (2003).

\bibitem{Ervin'PRB'2003} S. C. Ervin and I. I. Mazin,
Phys. Rev. B \textbf{68}, 132505 (2003).

\bibitem{Gonnelli'PRB'2005} R. S. Gonnelli, D. Daghero, A. Calzolari,
G. A. Ummarino, V. Dellarocca, V. A. Stepanov, S. M. Kazakov, N.
Zhigadlo, and J. Karpinski, Phys. Rev. B \textbf{71}, 060503
(2005).

\bibitem{Sologubenko'PRB'2005} A. V. Sologubenko, N. D. Zhigadlo,
S. M. Kazakov, J. Karpinski, and H. R. Ott, Phys. Rev. B
\textbf{71}, 020501 (2005).

\bibitem{Pallecchi'cond-mat'0508509} I. Pallecchi, V. Ferrando,
E. Galleani D'Agliano, D. Marr\'{e}, M. Monni, M. Putti, C.
Tarantini, F. Gatti, H. U. Aebersold, E. Lehmann, X. X. Xi, E. G.
Haanappel, and C. Ferdeghini, cond-mat/0508509 (unpublished).

\bibitem{Kogan'PRB'2004} V. G. Kogan and N. V. Zhelezina,
Phys. Rev. B \textbf{69}, 132506 (2004).

\bibitem{Carrington'cond-mat'0504664} A. Carrington, J. D. Fletcher,
J. R. Cooper, O. J. Taylor, L. Balicas, N. D. Zhigadlo, S. M.
Kazakov, J. Karpinski, J. P. H. Charmant, and J. Kortus, Phys.
Rev. B \textbf{72}, 060507(R) (2005).

\bibitem{Bussmann-Holder'PRB'2003} A. Bussmann-Holder and A.
Bianconi, Phys. Rev. B \textbf{67}, 132509 (2003).

\bibitem{Ummarino'PhysicaC'2004} G. A. Ummarino, R. S. Gonnelli,
S. Massidda, and A. Bianconi, Physica C \textbf{407}, 121 (2004).

\bibitem{Cooley'cond-mat'0504463} L. D. Cooley, A. J. Zambano,
A. R. Moodenbaugh, R. F. Klie, Jin-Cheng Zheng, and Yimei Zhu,
cond-mat/0504463 (unpublished).

\bibitem{Ummarino'PRB'2005} G. A. Ummarino, D. Daghero, R. S.
Gonnelli, and A. H. Moudden, Phys. Rev. B \textbf{71}, 134511
(2005).

\end{thebibliography}
\end{document}